\pdfoutput=1

\documentclass{article}

\usepackage[nonatbib,final]{nips_2018}

\usepackage[utf8]{inputenc} 
\usepackage[T1]{fontenc}    
\usepackage{url}            
\usepackage{booktabs}       
\usepackage{amsfonts}       
\usepackage{nicefrac}       
\usepackage{microtype}      
\usepackage{textcomp}
\usepackage{amsmath}
\usepackage{graphicx}
\usepackage{multirow}
\usepackage{multicol,color}
\usepackage{hyperref}       

\title{Secure Computation for Machine Learning With SPDZ}

\author{
  Valerie Chen\\
  Yale University\\
  \texttt{v.chen@yale.edu} \\
  \And
  Valerio Pastro\\
  Yale University\\
  \texttt{valerio.pastro@yale.edu}\\
  \And
  Mariana Raykova \\
  Yale University \\
  \texttt{mariana.raykova@yale.edu} \\
}

\begin{document}

\maketitle

\begin{abstract}
Secure Multi-Party Computation (MPC) is an area of cryptography that enables computation on sensitive data from multiple sources while maintaining privacy guarantees.
However, theoretical MPC protocols often do not scale efficiently to real-world data. This project investigates the efficiency of the SPDZ framework~\cite{SPDZ12}, which provides an implementation
of an MPC protocol with malicious security, in the context of popular machine learning (ML) algorithms. In particular, we chose applications such as linear regression and logistic regression,
which have been implemented and evaluated using semi-honest MPC techniques~\cite{GSB0DZE17, MZ17}. We demonstrate that the SPDZ framework outperforms these previous implementations while providing stronger security.
\end{abstract}

\section{Introduction}

Many machine learning techniques, including regression analysis, aim to build a model that fits a set of predictors to a dependent variable. Such techniques are widely used to model and analyze big data.
In many settings, however, the input data for such ML analysis tools is partitioned among different parties, which have strict privacy policies. For example, if the Center for Disease Control is interested in identifying disease outbreak, they might want to incorporate patient data from many individual hospitals. The problem is that openly sharing this data for prediction or model-building purposes is against modern day privacy laws as it would leak private individual data. This is one of many real-world applications that could benefit from MPC, which allows parties to evaluate the output of the analysis without revealing more about the private inputs.

\subsection{Multi-Party Computation}

Secure MPC addresses the above problem by providing a mechanism through which different parties can run a joint computation over their private inputs with guarantees that the only thing
revealed about the inputs is the output of the computation and whatever can be inherently inferred from it.
There are two main types of MPC protocols in terms of their security guarantees: semi-honest and malicious protocols \cite{Oded:2009:FCV:1804390, Hazay10anote}. In semi-honest security, it is assumed that the parties will follow the protocol as specified, but they can try to infer information about the input from the protocol messages. In malicious security, dishonest parties may attempt to deviate from the specified protocol, and the protocol must guarantee that these parties cannot learn about the inputs. Since malicious protocols have to satisfy stronger guarantees in general, such construction are less efficient than semi-honest protocols.

Two recent works~\cite{GSB0DZE17, MZ17} propose efficient implementation of several central machine learning building blocks including conjugate gradient decent (CGD) and stochastic gradient descent (SGD) as well as their applications in linear and logistic regression. The work of Gascon et al.~\cite{GSB0DZE17} uses the framework for semi-honest computation Obliv-C~\cite{zahur2015obliv} and proposes
several different methods for solving systems of linear equations. Their main premise is to use an iterative method such as CGD and demonstrate trade-offs that saves computation and hence efficiency for the MPC in
return for a small accuracy loss. They further propose a modification of CGD that has stable behavior using fixed-point arithmetic because emulating floating point with the underlying MPC representation introduces substantial
efficiency overhead.

The work of Mohassel and Zhang~\cite{MZ17} utilizes stochastic gradient descent as a method for linear regression and logistic regression with the incorporation of different activation functions. The authors consider
arithmetic representation for the computation and propose new secure computation techniques for matrix computation, which generalizes the approach for generating multiplicative triples in a preprocessing step.
 Similarly to the work of Gascon et al.~\cite{GSB0DZE17}, this paper considers techniques for approximation that save computation. For example, they use a piece-wise approximation of the logistic function. The authors
 also propose new techniques for more efficient approximate computation of fixed-point encodings.

The techniques in both of the above works are restricted to the setting of two party computation. We selected the SPDZ~\cite{SPDZ12,KPR18,KSS13} framework for our experiments since it is one of the main and most comprehensive 
implementations for multiparty computation protocols, which provide malicious security and support more than two parties.

\section{ML Functionalities}

For our implementation we consider the same algorithms and functionalities as in the above two papers. Next we provide a brief overview of these classic algorithms (details can be found in ~\cite{HTF09}).



\subsection{Direct vs. Iterative Decomposition for Solving a Linear System}

Solving a system of linear equations, which underlies linear regression learning, can be done using techniques for direct and indirect decomposition.
LDLT and Cholesky are both variants of direct decomposition methods which decompose a Hermitian, positive-definite matrix into a lower triangular matrix and its conjugate transpose. The algorithms are cubic in complexity with asymptotic run time of $O(d^3)$, where $d$ is the dimension of the input matrix. 
The difference between LDLT and Cholesky is that Cholesky requires a square root. The representation of square root computation as an arithmetic circuit used in the MPC computation in SPDZ introduces 
considerable overhead. That is why we used the iterative Newton method as a way of approximating the square root computation. It computes $x_i$, where $x_{i}^2 = S$, with repetition of the following
update function $x_{n+1} =\frac{1}{2}(x_{n} + S/x_{n})$.

In terms of an iterative approach to regression, we used the approach proposed by
Gascon et al.~\cite{GSB0DZE17}, which uses a normalized version of CGD that preserves stability and convergence rate with fixed-point number representation. Similarly to other MPC implementations
using floating point representation in SPDZ introduces substantial efficiency overhead. 


\subsection{Stochastic Gradient Descent}

Stochastic gradient descent is an iterative approximation method that converges to the global minimum for convex problems, like linear and logistic regression. It is also a driving mechanism for non-convex problems like neural networks. An SGD iteration updates a weight vector $\mathbf{w}$ using a randomly selected sample from the training input as follows: $w_j := w_j - \alpha ({\partial C_i(\mathbf{w})}/{\partial w_j})$ with learning rate $\alpha$. In this update $C_i$ is the cost function, which can be instantiated with different concrete functions to obtain computation for linear regression and logistic regression. A common technique for SGD
computation is called \emph{mini-batch} -- instead of selecting one sample per iteration, a small batch of size $B$ samples are selected and the update function is performed averaging the partial derivatives across all samples. We use the mini-batch SGD in our implementation to obtain accuracy benefits. While the work of Mohassel and Zhang~\cite{MZ17} has optimizations for matrix computation, which can be used with mini-batch, for 
SPDZ this does not lead to additional savings.

\subsubsection{Linear and Non-Linear Activation Functions}

To obtain a solution for linear regression using SGD, we instantiate the update function of a learned weight as $w_j = w_j  - \alpha(X_i \cdot w *-y_i)X_{ij}$, where $X$ is the input matrix and $y$ is the input vector. In this update function, the weights are adjusted element-wise by the error from the predicted and expected value at a rate determined by $\alpha$.

Logistic regression is a classification algorithm for modeling a binary dependent variable. Logistic regression fits the logistic function $f(u) = \frac{1}{1+e^-u}$ to the input. The corresponding update function for mini-batched SGD for logistic regression is $w = w - \frac{1}{|B|} \alpha X^{T}_{B} \times f(X_{B} \times w - Y_{B})$, where $f$ maps the predicted value into the binary output space. Mohassel and Zhang [8] proposed the following piecewise function as approximation for $f$:

\[
  f(u) =
  \begin{cases}
                                   0 & \text{if $u < -0.5$} \\
                                   u + 0.5 & \text{if $-0.5 \leq u \leq 0.5$} \\
1 & \text{if $u > 0.5$}
  \end{cases}
\]

We compare the results of this MPC-friendly piecewise function to a more standard approximation approach of taking the Taylor Series expansion to varying degrees.

\section{Experiments}

\subsection{Experimental Setup}

For our evaluation, we implemented all algorithms both in the SPDZ framework as well as in python as a plaintext verification of the algorithm.
The main metrics of evaluations were the latency of the MPC computation and the accuracy error, and we aimed to explore the trade-offs between accuracy and efficiency. We varied the precision after the decimal point 
depending on what was used in the works that we compared against (32 and 64 bits for the linear regression, less for SGD).

We evaluated our methods on both real-world datasets (MNIST, Arcene, and 9 other UCI open-source datasets) as well as synthetically generated data. These real-world datasets allow us to compare the accuracy results to existing works and to demonstrate that SPDZ can be used in practical settings. We used synthetic data in order to explore larger ranges of data characteristics such as dimension ($d$ = 10, 20, 50, 100, 200, 500), condition number ($cd$ = [1,10]), and number of examples ($n$ = 1000, 100000).

Most of our experiments were ran using machines on the same local area network where there is no network latency. We performed tests where both parties were deployed on separate Amazon EC2 m4.large instances (see Figure~\ref{fig:result4}).
We also ran experiment with up to four parties (see Figure~\ref{fig:result3}).


\subsection{Results}

In this section we present empirical results for our SPDZ implementations evaluated with real and synthetically generated databases. We compare the five different algorithms in terms of accuracy and run time for various parameters. 

\begin{figure}[h!]
\centering
  \includegraphics[scale=0.4]{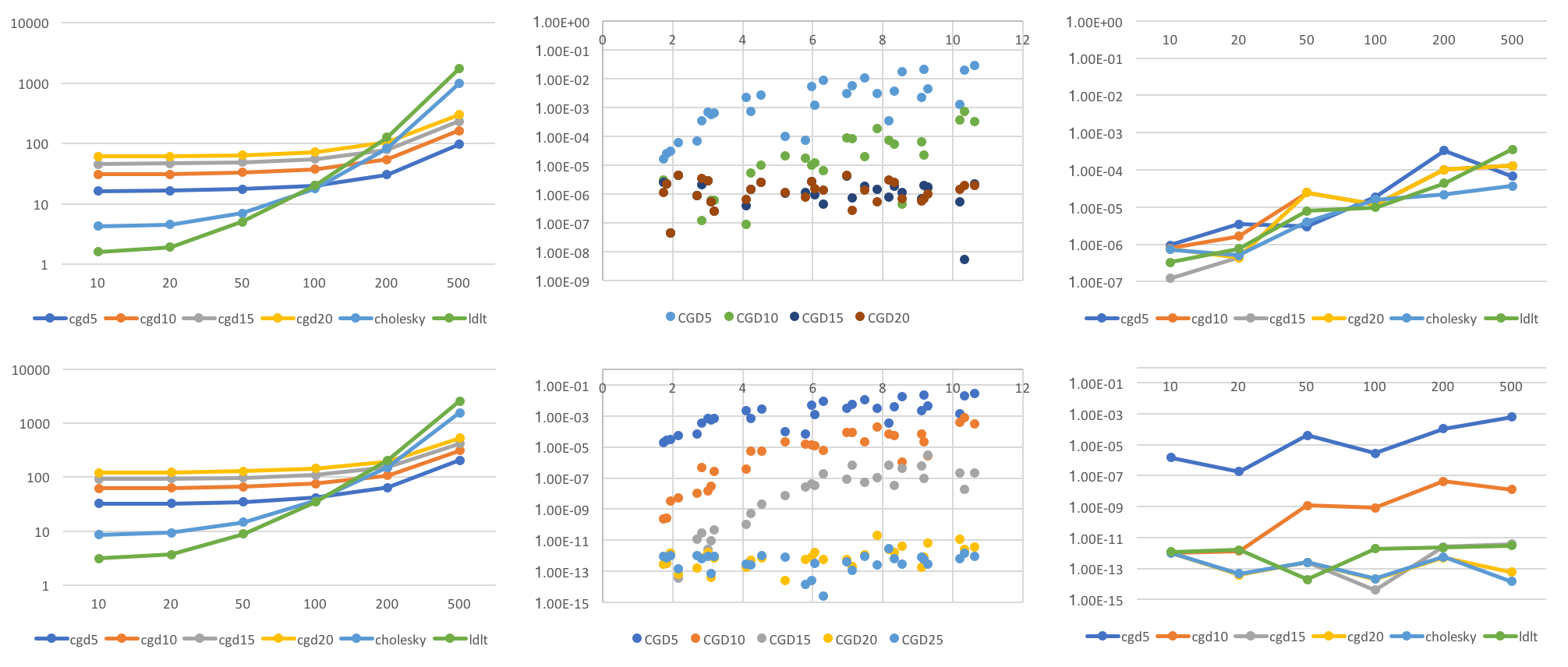}
    \caption{(Left) Run time as a function of input dimension. (Middle) Condition number as a function of accuracy. (Right) Accuracy as a function of the input dimension. (Top) Fixed-point with 60 bits of precision. (Bottom) Fixed-point with 28 bits of precision}
   \label{fig:result1}   
\end{figure}

For LDLT, Cholesky and various iterations of CGD, we evaluated on synthetically generated data of varying sizes and condition numbers. The larger the condition number is, the larger the error in approximations of the solution is. The direct decomposition methods grew exponentially in run time as input size increases, which is shown in the left column of Figure~\ref{fig:result1} -- this unlikely to be suitable for large size real data. Alternatively, the iterative CGD runtime increases at a much slower rate. In the middle column of Figure 2, we find that about 20 iterations are sufficient to reach maximum accuracy given the number of allocated bits even with varying condition numbers. Particularly for the 64-bit case, shown on the bottom right, the accuracy is identical for CGD after 15 iterations and Cholesky/LDLT.

\begin{figure}[h!]
\vspace{-4mm}
\centering
  \includegraphics[scale=0.6]{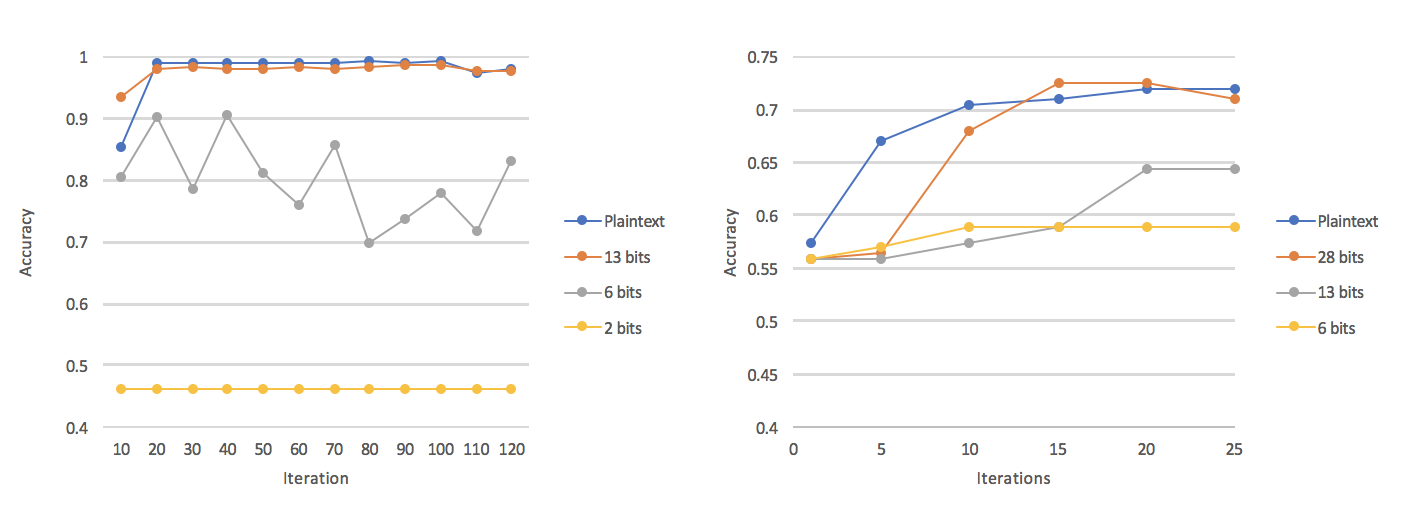}
  \vspace{-4mm}
   \caption{Comparing accuracy of privacy preserving linear regression with various fixed point precisions and plaintext training on floating point for MNIST (left) and Arcene (right).}
   \label{fig:result2}  
\end{figure}

Figure~\ref{fig:result2} compares SGD on MNIST and Arcene results. It shows that the number of bits of precision needed to get good accuracy is highly dependent on the dataset. For MNIST, 13 bits was sufficient to match plaintext accuracy, but 28 bits were needed for Arcene. The MNIST data contains only 784 features while there are 10,000 in the Arcene data, 3,000 of which are considered "probes" with no predictive power, which could explain the lower overall accuracy of \cite{MZ17}. While the numbers in the MNIST data ranged from 0 to 9, Mohassel and Zhang~\cite{MZ17} only used 0s and 1s labels from the dataset, reducing it to a binary problem. We replicated this approach and present the results below. We did run the computation to predict all 10 digits, but found that SGD only achieved a much lower accuracy of about 19\%. We also compared the root mean squared error (RMSE) of SGD on 9 UCI open-sourced datasets of ranging sizes to results in \cite{GSB0DZE17}. Our results in the SPDZ secure setting typically increased RMSE by about $5-20\%$ compared to plaintext computation, but still outperformed RMSE results from \cite{GSB0DZE17} in both CGD and SGD.

In terms of logistic regression, for SPDZ, we did not find that the new activation function was a better alternative to taking a Taylor Series approximation for the exponential function as shown in Table~\ref{tbl1}. We found that for SPDZ, which is based on arithmetic circuits, the extra time to take a few extra degrees in the approximation was negligible.  

\begin{table}[h!] 
\caption{Comparing the validation accuracy for different activation functions for logistic regression.}
\centering
\label{my-label}
\begin{tabular}{@{}lllllll@{}}
\toprule
       & Plaintext & New Activation function & \multicolumn{4}{l}{Polynomial Approximation} \\
       &           &                         & degree 2  & degree 5  & degree 7 & degree 10 \\ \midrule
MNIST  & 99.9\%    & 95\%                    & 97\%      & 85\%      & 91\%     & 99.5\%    \\
Arcene & 72.0\%    &           44.0\%              &   44.0\%        &    44.5\%       &    65\%      &    72\%       \\ \bottomrule
\end{tabular}
\vspace{-4mm}
\label{tbl1}
\end{table}

\subsection{Conclusion}

SPDZ was able to achieve comparable accuracy for LDLT, Cholesky, and CGD when compared to Obliv-C with runtime faster by an order of magnitude on larger matrix sizes even in a distributed machine setting. SPDZ also achieved lower RMSE than Obliv-C using SGD. SPDZ was able to match accuracy and latency results for SecureML on SGD and Logistic Regression. This result is promising for SPDZ to be extended to more complex algorithms including neural networks with hidden layers.


\bibliography{nips_format}

\begin{thebibliography}{1}

\bibitem{SPDZ12}
Ivan Damg{\aa}rd, Valerio Pastro, Nigel~P. Smart, and Sarah Zakarias.
\newblock Multiparty computation from somewhat homomorphic encryption.
\newblock In {\em Advances in Cryptology - {CRYPTO} 2012 - 32nd Annual
  Cryptology Conference, Santa Barbara, CA, USA, August 19-23, 2012.
  Proceedings}, pages 643--662, 2012.

\bibitem{GSB0DZE17}
Adri{\`{a}} Gasc{\'{o}}n, Phillipp Schoppmann, Borja Balle, Mariana Raykova,
  Jack Doerner, Samee Zahur, and David Evans.
\newblock Privacy-preserving distributed linear regression on high-dimensional
  data.
\newblock {\em PoPETs}, 2017(4):345--364, 2017.

\bibitem{HTF09}
Trevor Hastie, Robert Tibshirani, and Jerome~H. Friedman.
\newblock {\em The elements of statistical learning: data mining, inference,
  and prediction, 2nd Edition}.
\newblock Springer series in statistics. Springer, 2009.

\bibitem{Hazay10anote}
Carmit Hazay and Yehuda Lindell.
\newblock A note on the relation between the definitions of security for
  semi-honest and malicious adversaries ?, 2010.

\bibitem{KPR18}
Marcel Keller, Valerio Pastro, and Dragos Rotaru.
\newblock Overdrive: Making {SPDZ} great again.
\newblock In {\em Advances in Cryptology - {EUROCRYPT} 2018 - 37th Annual
  International Conference on the Theory and Applications of Cryptographic
  Techniques, Tel Aviv, Israel, April 29 - May 3, 2018 Proceedings, Part
  {III}}, pages 158--189, 2018.

\bibitem{KSS13}
Marcel Keller, Peter Scholl, and Nigel~P. Smart.
\newblock An architecture for practical actively secure mpc with dishonest
  majority.
\newblock In {\em Proceedings of the 2013 ACM SIGSAC Conference on Computer
  \&\#38; Communications Security}, CCS '13, 2013.

\bibitem{MZ17}
Payman Mohassel and Yupeng Zhang.
\newblock Secureml: {A} system for scalable privacy-preserving machine
  learning.
\newblock In {\em 2017 {IEEE} Symposium on Security and Privacy, {SP} 2017, San
  Jose, CA, USA, May 22-26, 2017}, pages 19--38, 2017.

\bibitem{Oded:2009:FCV:1804390}
Goldreich Oded.
\newblock {\em Foundations of Cryptography: Volume 2, Basic Applications}.
\newblock Cambridge University Press, New York, NY, USA, 1st edition, 2009.

\bibitem{zahur2015obliv}
Samee Zahur and David Evans.
\newblock Obliv-c: A language for extensible data-oblivious computation.
\newblock Cryptology ePrint Archive, Report 2015/1153, 2015.
\newblock \url{https://eprint.iacr.org/2015/1153}.

\end{thebibliography}
\bibliographystyle{plain}

\newpage 

\section*{Supplementary Material}

\subsection{Three and four party runtimes}

In addition to the 2 player case, we also ran experiments with 3 and 4 players on the local network, where the input was vertically partitioned between parties.

\begin{figure}[h]
\centering
  \caption{Run time results for 3 and 4 players in 32 bits. (Left) 3 Players. (Right) 4 Players. }

  \includegraphics[scale=0.6]{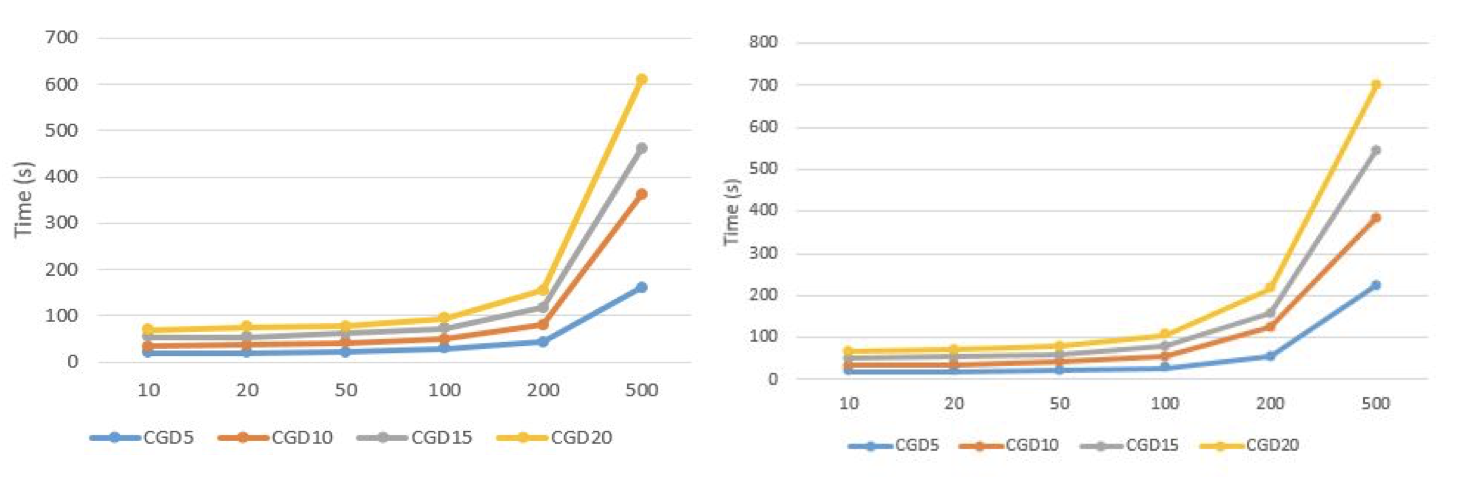}
  \label{fig:result3}
\end{figure}

\subsection{EC2 runtimes}

We deployed 2 AWS EC2 Instances that ran the SPDZ protocol. The EC2 runtime results are comparable to the local network results for smaller matrix sizes but increases at a faster rate for larger matrix sizes.

\begin{figure}[h]
\centering
  \caption{Run time results for 2 players deployed on EC2 instances averaged over 15 runs. (Left) 32 bits. (Right) 64 bits.}
  \includegraphics[scale=0.7]{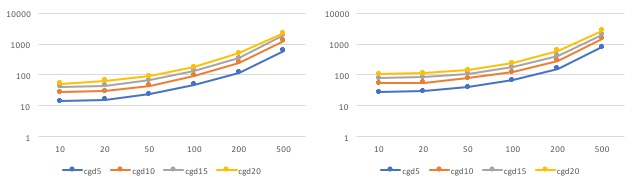}
  \label{fig:result4}
\end{figure}

\subsection{Activation function runtime comparison}

To determine the difference in efficiency in the two activation functions, the new one proposed by ~\cite{MZ17} and the other method which they claimed to be inefficient, we ran results for SPDZ on two different datasets for a few different variations of the activation functions. The compile and run times correspond to the offline and online phase for SPDZ respectively.

\begin{table}[h]
\centering
\label{my-label1}
\caption{Comparison of efficiency for the proposed activation function in ~\cite{MZ17} and Taylor Series approximation in SPDZ.}
\begin{tabular}{@{}lcccc@{}}
\toprule
                        & \multicolumn{2}{c}{MNIST} & \multicolumn{2}{c}{Arcene} \\ \midrule
                        & Compile      & Run        & Compile      & Run         \\
New Activation Function & 26.01        & 18.8       & 220.14       & 224.47      \\
2 Polynomial            & 27.79        & 19.28      & 222.85       & 239.98      \\
5 Polynomial            & 30.32        & 19.8       & 225.33       & 241.59      \\
7 Polynomial            & 31.99        & 20.65      & 228.33       & 239.11      \\
10 Polynomial           & 34.24        & 20.71      & 230.81       & 240.79      \\ \bottomrule
\end{tabular}
\label{tbl2}
\end{table}

\newpage
\subsection{RMSE results}

SGD was also evaluated on 9 different datasets selected from the UCI repository. Further details about the specific datasets, including the regularization parameter used, are detailed in ~\cite{GSB0DZE17}. The dimensions of the problems ranged from 7 to 384 with the number of examples ranging from over 200 to almost 3 million.

\begin{table}[h]
\centering
\label{my-label2}
\caption{SGD Results for 9 open sourced UCI datasets.}
\newcommand*{\MyIndent}{\hspace*{1.2cm}}%
\begin{tabular}{@{}llllll@{}}
\toprule
Dataset Name          & \multicolumn{1}{c}{SGD Plaintext} & \multicolumn{2}{c}{\begin{tabular}[c]{@{}c@{}}SPDZ (28 bit)\\ Error \MyIndent Time\end{tabular}} & \multicolumn{2}{c}{\begin{tabular}[c]{@{}c@{}}SPDZ (13 bit)\\ Error \MyIndent Time\end{tabular}} \\ \midrule
Student Performance   & 0.11                              & 0.12 (+8.34\%)                                  & 2.174                                  & 0.12(+8.34\%)                                   & 2.616                                  \\
Auto MPG              & 0.56                              & 0.68 (+21.4\%)                                  & 0.663                                  & 0.68(+21.4\%)                                   & 0.704                                  \\
Communities and Crime & 0.06                              & 0.17 (+183\%)                                   & 10.352                                 & 0.19 (+216\%)                                   & 12.363                                 \\
Wine Quality          & 0.18                              & 0.19 (+5.55\%)                                  & 0.809                                  & 0.19 (+5.55\%)                                  & 0.969                                  \\
Bike Sharing          & 0.23                              & 0.24 (+4.34\%)                                  & 0.910                                  & 0.25 (+8.69\%)                                  & 1.097                                  \\
Blog Feedback   & 0.04                              & 0.04 (+0.0\%)                                   & 9.455                                  & 0.04 (+0.0\%)                                   & 9.091                                  \\
CT Slices   & 0.22                              & 0.22 (+0.0\%)                                   & 16.458                                  & 0.22 (+0.0\%)                                   & 16.295                                  \\
Year Prediction MSD   & 0.06                              & 0.06 (+0.0\%)                                   & 7.146                                  & 0.06 (+0.0\%)                                   & 8.513                                  \\
Gas Sensor Array      & 0.20                              & 0.36 (+20.0\%)                                  & 1.151                                  & 0.36 (+20.0\%)                                  & 1.559                                  \\ \bottomrule
\end{tabular}\\
\label{tbl3}
\end{table}

\end{document}